\documentclass[a4paper,11pt]{article}
\usepackage{pos}
\renewcommand{\printHeadAuthors}{Gunnar Bali et al.}
\title{Lattice gauge ensembles and data management}
\author[a,\#]{Gunnar Bali}
\affiliation[a]{Institut für Theoretische Physik, Universität Regensburg,
  Universitätsstr.~31, 93053 Regensburg, Germany}
\author*[b,1]{Ryan~Bignell}
\affiliation[b]{Department of Physics, Swansea University,
  Swansea SA2 8PP, United Kingdom}
\author*[c,2]{Anthony~Francis}
\affiliation[c]{Institute of Physics, National Yang Ming Chiao Tung University,
  30010 Hsinchu, Taiwan}
\author*[d,3]{Steven~Gottlieb}
\affiliation[d]{Department of Physics, Indiana University,
  Bloomington, IN 47405, USA}
\author*[e,4]{Rajan~Gupta}
\affiliation[e]{Los Alamos National Laboratory, Theoretical Division T-2,
  Los Alamos, NM 87545, USA}
\author*[f,5]{Issaku~Kanamori}
\affiliation[f]{RIKEN Center for Computational Science (R-CCS),\\
7-1-26, Minatojima Minamimachi, Kobe 650-0047, Japan}
\author*[g,6]{Bartosz~Kostrzewa}
\affiliation[g]{High Performance Computing and Analytics Lab, Rheinische Friedrich-Wilhelms-Universität Bonn,\\
Friedrich-Hirzebruch-Allee 8, 53115 Bonn, Germany}
\author*[h,7]{Andrey~Yu.~Kotov}
\affiliation[h]{Jülich Supercomputing Centre (JSC), Forschungszentrum Jülich,
  52428 Jülich, Germany}
\author*[i,8]{Yoshinobu~Kuramashi}
\affiliation[i]{Center for Computational Sciences, University of Tsukuba,
 305-8577, Tsukuba, Japan}
\author*[j,9]{Robert~Mawhinney}
\affiliation[j]{Physics Department, Columbia University,
  New York, NY 10027, USA}
\author*[k,10]{Christian~Schmidt}
\affiliation[k]{Fakultät für Physik, Universität Bielefeld,
  Universitätsstr.~25, 33615 Bielefeld, Germany}
\author*[a,11]{Wolfgang~Söldner}
\author*[l,m,12]{Peng~Sun}
\affiliation[l]{Institute of Modern Physics, Chinese Academy of Sciences,
  Lanzhou 730000, China}
\affiliation[m]{University of Chinese Academy of Sciences,
  Beijing 100049, China}
\note{For the FASTSUM Collaboration.}
\note{For the Open Lattice Initiative.}
\note{For the MILC Collaboration.}
\note{For the JLab, College of William \& Mary, LANL, MIT, OLCF, Marseille Collaboration.}
\note{For the JLQCD Collaboration.}
\note{For the Extended Twisted Mass Collaboration (ETMC).}
\note{For the TWEXT Collaboration.}
\note{For the PACS Collaboration.}
\note{For the RBC-UKQCD Collaborations.}
\note{For the HotQCD Collaboration.}
\note{For the CLS Community Effort.}
\note{For the CLQCD Collaboration.}
\note[\#]{Convenor}

\abstract{
  The generation of ensembles of gauge configurations is a
  considerable expense.
  The preservation and curation of these ensembles constitutes a
  valuable shared resource for the lattice field theory community.
  The organizers of Lattice 2022 dedicated a parallel session
  to the presentation of gauge ensembles and their generation,
  plans for ensemble publication and
  data management/storage activities of different
  collaborations. A summary of the twelve contributions is
  presented here.}
  
\FullConference{
The 39th International Symposium on Lattice Field Theory,\\
8th-13th August, 2022,\\
Rheinische Friedrich-Wilhelms-Universität Bonn, Bonn, Germany
}
\tableofcontents
\begin{document}
\maketitle
\section{Introduction}
Lattice QCD calculations factorize into at least three parts,
that we label ``generation,'' ``measurements'' and ``analysis.''
First of all, ensembles of gauge configurations are generated using
the Markov chain Monte Carlo method. Second, a set of correlation
functions that is relevant for the specific physics programme
is computed on these ensembles. This process is often described as
``making measurements'' on gauge configurations. Finally, the
correlation functions are statistically analysed and
combined into the observables of interest.
These data are then extrapolated to the infinite volume and
continuum limits, as well as to the quark masses of interest,
quantifying systematic and statistical errors.

The first two steps (generation and measurement) are compute intensive.
During measurements, the building blocks are quark propagators
with different types of sources. These are mostly computed on the fly and
not stored for re-use, while perambulators~\cite{HadronSpectrum:2009krc}
and eigenvectors, that have broader applications, are stored. The resulting
correlation functions are observable specific. Perambulators, eigenvectors and
correlation functions are mostly shared within one collaboration
or project. This may change in the future.
Gauge ensembles are much less specific, can easily be transferred
between computer centres and are often shared far beyond the group that
created them.

This parallel session is directed towards summarizing what gauge ensembles
exist and the respective data sharing policies and data management strategies.
We shall distinguish between ``data providers'' who generate gauge
ensembles and manage the data and ``data consumers'' who use these
(and can also be
providers). Measurements
with a specific physics goal typically require less computational
resources, planning and parameter tuning than a campaign to
generate a set of gauge ensembles. Therefore, the former sometimes
can be carried out by
smaller groups, without multiple computer time proposals and with less
coordination and administrative overhead. Apart from these accessibility
issues, gauge ensembles are highly valuable research products
and, once these exist, the science output should be maximized to further
our understanding of Nature.

Already twenty years ago, the lattice community initiated
the International Lattice Data Grid (ILDG)~\cite{Davies:2002mu,Irving:2008uk,Maynard:2009szr,Beckett:2009cb} as a framework to enable and facilitate
world-wide sharing of gauge configurations. This already
anticipated most of the FAIR (Findable, Acessible, Interoperable,
Reusable) principles~\cite{Wilkinson2016}. A major achievement of ILDG was the specification of a
community-wide agreed metadata schema (QCDml)~\cite{Coddington:2007gz} to
concisely markup the gauge configurations. ILDG is organized as a federation
of autonomous ``regional grids'' within a single Virtual Organization. It
has a uniform user registration and defines a common format for the binary
storage of the gauge configurations. Standardized interfaces for
the services, which are to be operated by each regional grid, like storage and a
searchable metadata catalogue, render these regional services interoperable.
The status and plans of the modernization of ILDG have been presented in a
plenary talk at this conference~\cite{Karsch:2022tqw}, as well as in a
lunch meeting, organized by the working groups of ILDG.

There exist different levels of the sharing of
gauge ensembles between the data providers and
the data consumers.
Sharing may be restricted to within a collaboration, e.g., between the
members who generate the configurations
and those who carry out measurements and the analysis.
Often, gauge ensembles which initially are only
shared internally are declared ``public'' after some embargo time,
with guidelines how these and associated publications should be cited.
Other levels of access and usage are possible, e.g., limited to
specific projects or sharing between
two or more collaborations.
In an ideal world, one may envisage the following scenario.
\begin{itemize}
  \setlength{\parsep}{0pt}
  \setlength{\itemsep}{0pt}
  \setlength{\topsep}{0pt}
\item In a {\bfseries collaboration-internal} context, the {\bfseries data consumers}
  know everything about the configurations being produced (availability and
  provenance, simulation parameters, storage location, etc.). All data
  management within the collaboration follows a clear and transparent
  data management plan and the usage rules are well-defined and
  known to everyone within the collaboration. The actual management of all
  the data and metadata is taken care of by designated members of the
  collaboration.
\item In a {\bfseries collaboration-internal} context, the
  {\bfseries data providers}
  follow a well-defined and smooth workflow
  for collecting and archiving all gauge configurations
  and the corresponding metadata. Internal and
  shared data can be handled in essentially the same way, so that
  data can be declared completely ``public'' or shared with specific
  external users with little extra effort (e.g., by requiring
  to switch an access permission flag for an ensemble at the end of an
  embargo time). Data that are declared ``public'' are
  easily citable, making their impact visible to
  funding agencies, computing centres and the wider research community.
\item In a {\bfseries community-wide} context, the {\bfseries data consumers}
  know
  about the existence of all publicly available gauge ensembles that
  may be relevant or interesting for their own research projects. All
  scientists in the community should obtain these at essentially
  no cost in terms of human or computing resources and then be able to
  freely carry out their high quality research. Every data consumer
  follows good scientific practice and properly acknowledges the source
  of the data and gives credit to the data providers.
\item In a {\bfseries community-wide} context, the {\bfseries data providers}
  can make their
  valuable data available on some storage infrastructure at no extra cost
  in terms of human or hardware resources. Declaring data ``public''
  will make these known to other researchers who will frequently use
  these in other projects, so that the data providers receive recognition and
  citations.
\end{itemize}
In the real world, where many of those responsible for generating, storing
and managing the data are on temporary
positions and where large, globally accessible, long-term storage
is not for free, the situation is more challenging.

In this contribution, we collect the
present status of ensemble generation to inform both data consumers
and providers
about the availability of gauge ensembles and present practices. We
restrict ourselves to simulations of QCD. At present,
these are mostly carried out
using $N_f=2+1$, $N_f=2+1+1$ and also $N_f=1+1+1+1$ sea quark flavours
$q=u, d, s, c$, with various fermion discretizations.
Naturally, we can only cover simulations by the groups who responded
to the call. The next section provides the current status.
This is followed by a brief summary.

\section{Contributions}
In this section, the different collaborations present their physics programmes,
gauge ensembles and data policies. The contributions are ordered
like the collaborations on the title page, i.e.,
alphabetically in terms of the
respective author names. The original presentations can be found
on the indico page of the conference~\cite{parallel}.

\subsection{FASTSUM Collaboration}
The FASTSUM collaboration uses $N_f=2+1$ flavour anisotropic gauge ensembles in the fixed-scale approach to study the behaviour of QCD in as a function of the temperature in the hadronic and plasma phases. Specifically, we have considered the behaviour of hadronic states including light, strange, charm and bottom quarks, the electrical conductivity of QCD matter, the interquark potential and properties of the chiral transition. The fixed-scale approach allows for a conceptually clear investigation of temperature effects while anisotropy allows more temporal points at each temperature.

FASTSUM gauge fields utilise an order $a^2$ improved Symanzik gauge action and an order $a$ improved spatially stout-smeared Wilson-Clover action following the parameter tuning and zero-temperature ensembles of the Hadron Spectrum collaboration~\cite{Edwards:2008ja,HadronSpectrum:2008xlg}. ``Generation 2'' ensembles were generated using the Chroma~\cite{Edwards:2004sx} software suite while the newer ``Generation 2L'' used a modfication~\cite{glesaaen_jonas_rylund_2018_2216355} of the {\sc openQCD}~\cite{Luscher:2012av,openqcd} package which introduces stout-link smearing and anisotropic actions. We use an anisotropy $\xi = a_s/a_\tau \sim 3.5$ with $a_s\sim 0.12$ fm, $N_s = 24$ or $32$ and a wide range of $N_\tau$ corresponding to $T \in[44,760]$ MeV. ``Generation 2'' and ``Generation 2L'' differ mainly in their quark mass. Full details of these ensembles may be found in ref.~\cite{Aarts:2014nba,Aarts:2020vyb}.

As we use a derivative of the {\sc openQCD} package log, files include information such as the {\sc openQCD} version, algorithmic parameters, plaquette values and the run time. Additionally we maintain a centralised metadata repository detailing (among other information) who was responsible for each run, on which machine that run was produced and where copies may be found. The gauge fields are redundantly stored on two (well-separated) storage servers managed by Swansea University in the openQCD format. The ``Generation 2'' ensembles are available upon request while other ensembles will be available after an embargo time. We anticipate making ensembles available through the next incarnation of ILDG with supplementary information also available on Zenodo~\cite{zenodo}.

\subsection{Open Lattice Initiative}
The increase of numerical cost and data complexity in precision era lattice QCD calculations poses new questions on data generation, transfer and access. 
The Open Lattice Initiative (OpenLat) was established as a response to these questions in 2019. The goal of this young initiative is to generate and make available QCD gauge ensembles for a broad range of physics applications. Emphasis is put on open science and user access.

The initiative uses the stabilised Wilson fermion (SWF) framework~\cite{Francis:2019muy}. An important new component is the exponentiated clover term in the fermion action, which cures a large volume pathology in Wilson-Clover fermions and exhibits benefits in terms of discretization effects~\cite{Francis:2022hyr}.
Furthermore SWFs include a number of algorithmic improvements, such as choosing the Stochastic Molecular Dynamics (SMD) algorithm, which shows favorable autocorrelation times and stability~\cite{Luscher:2011kk,Luscher:2017cjh}. Precision losses are further prevented by using a volume-independent solver stopping criterion and quadruple precision arithmetic in global sums.
These measures are implemented in addition to the established techniques, such as the Schwarz-alternating-procedure (SAP), local deflation, mass-preconditioning, multiple time-scale integrators and others, for details see ref.~\cite{Francis:2019muy,Francis:2022hyr}. For generation we use the open source software packages \texttt{openQCD-2.0} and \texttt{2.4}~\cite{openqcd}.

All generated gauge ensembles are $N_f=2+1$ full QCD. The scale is determined using the gradient flow time criterion~\cite{Luscher:2010iy} and results are converted to physical units via $\sqrt{8t_0}=0.414(5)\textrm{fm}$~\cite{Bruno:2016plf}.
By default (anti-)periodic boundary conditions are set with the aim to preserve translation invariance. Open boundaries are chosen once a significant slowdown  in topological tunneling is observed.

The production plan of ensembles proceeds in three stages~\cite{Cuteri:2022gmg}: 
In stage 1, after a high precision tuning, ensembles are generated at the $SU(3)$-flavor symmetric point ($m_\pi=m_K=412$~MeV) at four lattice spacings, $a=0.12, 0.094, 0.077$ and $0.064$~fm. 
In stage 2, the light quark masses are reduced to $M_\pi=300$ and $200$~MeV,  keeping the quark mass matrix trace fixed, and the lattice spacing $a=0.055$~fm (open boundary conditions) is added.
In stage 3, the light quark masses are lowered to the physical point.
Throughout $L\gtrsim 3$fm with $m_\pi L\gtrsim 4$ and $T/L\geq2$.
The initiative is currently in the process of completing stage 1.
Each completed stage is accompanied by a reference publication. With this publication all configurations and metadata are made openly available without further embargo time. The public data are: the metadata catalogue, the gauge configurations and the basic observables, see ref.~\cite{Francis:2022hyr} for a list.
The metadata catalogue follows a detailed provenance policy and will be compliant with community standards, such as ILDG. All metadata is preserved on disk and in the main online repository~\cite{openlat}. 
Users may obtain access to the configurations of ongoing, i.e. unpublished, stages. This user-access is granted on a case-by-case basis. 

The initiative maintains a repository of 100~TB total, which is projected to grow to 500+ TB with the completion of stage 2. At this point the repository includes $\sim 20$k saved configurations in the so-called openQCD format. The data is mirrored at two separate locations on disk and tape. 
As per openQCD format standard the plaquette is given in the header of each configuration file, additionally \verb+sha512+ checksums are kept separately as means to monitor data integrity.

\subsection{MILC Collaboration}
The MILC Collaboration has been sharing code and lattice configurations for
approximately 30~years. We started with $N_f=2$ staggered quarks, transitioned
to the asqtad action with $N_f=2+1$, and are now using the 
HISQ action~\cite{Follana:2006rc} with
$N_f=2+1+1$ or $1+1+1+1$.  Our original physics interests included both zero
and non-zero temperature QCD. Our work on non-zero temperature QCD culminated
with our participation in HotQCD, which is covered later in this article. Our
main interests now are leptonic and semileptonic decays of mesons containing
charm or bottom quarks; properties of baryons using staggered quarks,
all currently done together with the Fermilab Lattice Collaboration;
and the muon anomalous magnetic moment, done with the Fermilab
Lattice and HPQCD Collaborations.  
We are also interested in electromagnetic corrections.

We were initial users of the Gauge Connection at NERSC where many asqtad
ensembles were made available in the ILDG format. Unfortunately, this service is
in disarray, but there are plans for its revival. The asqtad physics program
and the associated ensembles were extensively reviewed in 
ref.~\cite{MILC:2009mpl}. The more modern ensembles using the HISQ action
are our current focus. We have approximately 25 publicly available ensembles
with lattice spacings $a\approx 0.15$, 0.12, 0.09, 0.06, and 0.042~fm.  
At each lattice spacing, there is one ensemble where the 
light, strange, and charm
quark masses are closely tuned to physical values. There are also
several ensembles with the light quark mass higher than its physical
value, enabling study of the chiral limit.
At 0.12~fm, with $m_l=0.1 m_s$, we have three volumes, $L=24$, 32, and 40.
In addition,
we have an ensemble with $a\approx 0.03$~fm with the light quark
set to 0.2
times the strange quark mass. With the advent of exascale computers, we hope
to generate a new ensemble with the light quark mass tuned so that the Goldstone
pion mass is 135~MeV. There are also a number of ensembles in which the
strange quark mass is reduced from its physical value to better enable chiral
fits and determination of the low energy constants in the chiral lagrangian.
It should also be mentioned that CalLat has been generating additional
HISQ ensembles~\cite{Miller:2020evg}.

Configurations are generated using the rational hybrid molecular dynamics or
rational hybrid Monte Carlo algorithms~\cite{Clark:2003na}. We also employ
the Hasenbusch method~\cite{Hasenbusch:2001ne}.   
Details may be found in refs.~\cite{MILC:2010pul} and~\cite{MILC:2012znn}.
The performance
of the MILC code, on various architectures, is enhanced by using
QOP~\cite{USQCD_SciDAC}, QPhiX~\cite{Joo:2013lwm,Li:2014kxa,DeTar:2016ndn,DeTar:2018pyj}, or
QUDA~\cite{Barros:2008rd,Clark:2009wm,Babich:2011np,QUDADownload}.

We have the log files from most of the gauge generation runs. These contain
checksums for the configurations, the plaquette, the chiral condensate, and many
Wilson loops. The configurations contain metadata including two checksums, 
the space-space and space-time plaquette, run parameters, and the time 
the configuration was generated. This information is also contained in an 
ASCII ``info'' file of about 600 bytes. Not all log files contain the name
of the machine on which the code was run, but it is usually possible to 
tell from other details in the log file. We are trying to be more systematic
about including such information, and the date on which the code was compiled.
Both asqtad and HISQ ensembles are archived
at Fermilab for the convenience of USQCD members. A second copy of most HISQ
ensembles is kept on Ranch at the Texas Advanced Computing Center (TACC).  
A few years ago, we had to migrate our data from one tape system to a 
new one, and it was decided not to preserve the asqtad ensembles at TACC.

Our sharing policy is available on GitHub under the milc-qcd main page.  
The information
is on the wiki of the repository called ``sharing''~\cite{MILCwiki}.
It contains links to our sharing policy, a web page
with a detailed list of the publicly available ensembles, and a document 
explaining which paper to cite when using each configuration and how we 
would like to be acknowledged.

\subsection{JLab, College of William \& Mary, LANL, MIT, OLCF, Marseille Collaboration}
This US community effort is generating ensembles with $2+1$ flavors of  Wilson-clover fermions with stout-link smearing
of the gauge fields and a tree-level tadpole-improved Symanzik gauge action. 
One iteration of the four-dimensional stout smearing is used with the weight $\rho = 0.125 $ for the staples in the rational hybrid Monte Carlo (RHMC) algorithm. After stout smearing, the tadpole-improved tree-level clover coefficient $C_{SW}$ is very close to the nonperturbative value. This was confirmed using the Schr\"odinger functional method for determining the clover coefficient nonperturbatively. The tuning of the strange quark mass has been done in two ways.  For the ensembles at $a = 0.127$ and $0.091$~fm, we have required the ratio $(2M_{K^+}^2 - M_\pi^2)/M_{\Omega^-}$ to take on its physical value $0.1678$. This tuning is done in the 3-flavor theory, and the resulting value of $m_s$  is then kept fixed as the light-quark masses in the (2+1)-flavor theory are decreased towards their physical values. For the other ensembles, we have 
required $M_K = 495$~MeV, independent of the light quark masses. The lattice scale has, so far, been set using $w_0$. 

In the HMC algorithm, the light quark generation is carried out utilizing the 
Hasenbusch method~\cite{Hasenbusch:2001ne} of factorizing the 2-flavor determinant with chains of 3--4 
ratios followed by a 2-flavor term. The Hasenbusch ratios are implemented by 
using different quark masses in the numerator and denominator of 
the determinant ratios, rather than by utilizing different twisted 
masses in these terms. The one flavor piece was carried out using a rational 
approximation for $\sqrt{M^{\dagger}_s M_s}$. 
The two flavor solves are carried out using the multi-grid preconditioned 
GCR solver implemented in QUDA~\cite{Clark:2016rdz,Clark:2009wm}. 

The integration is carried out with a nested Force Gradient integrator. We tuned the placement of terms in the Hamiltonian (monomials in the action) on various timescales, the number of steps on each timescale, and the Hasenbusch masses to minimize trajectory time while maintaining an acceptance rate of approximately 95\% for our 4th order integrator.  The procedure was somewhat ad-hoc, but primarily involved: a) tuning the quark masses so that the Hasenbusch ratio terms would have forces that were roughly equal in infinity-norm and less than 1 in value. Thereafter, we placed the rational approximation on the middle timescale and the gauge and 2-flavor fermion pieces on the finest timescale. Generally we set the step on the finest timescale to be very large and adjusted the middle time-scale to a level, where we would have no instability from the rational term having too big a time-step. Finally we reduced the number of steps on the finest time step.  

Beyond making the multi-grid blocking sizes fit, we did not perform extensive 
tuning of the multigrid parameters from ensemble to ensemble.  We refresh the 
multi-grid subspace through ‘polishing’ --- iterating the existing subspace vectors 
with the current operator, once a given threshold is exceeded. If after the polish 
the solution does not converge within the iteration limit we regenerate the subspace. 
If ever the solution fails to converge even after regenerating the subspace we terminate the program.

Our computations are carried out with the Chroma code~\cite{Edwards:2004sx,ChromaDownload} built 
over QDP-JIT~\cite{Winter:2014dka,QDPJITDownload} 
and QUDA~\cite{Clark:2009wm,Babich:2011np} and the majority of the 
ensembles have been generated on the Summit system at OLCF in a variety of allocations. 
So far 13 ensembles have been generated at four values 
of the lattice spacing ($a \approx 0.127, \ 0.091,\ 0.071$ and $0.055$~fm) and 
four values of the pion mass $M_\pi = 270,\  220,\ 170$ and $130$~MeV with 
between 10,000 and 20,000 thermalized trajectories each. At present 1000 
lattices, each separated by 4 trajectories, are available on request to the USQCD 
collaboration for non-competing physics projects and the full set will be made 
available by the beginning of 2024 through a password protected  repository at OLCF.

\subsection{JLQCD Collaboration}
The JLQCD collaboration puts emphasis on the chiral symmetry 
and uses M\"obius domain-wall fermions~\cite{Brower:2012vk}
with scale factor 2 (Shamir type).
Stout smearing with $\rho=0.1$ and $n=3$ is applied to the fermion action.
The tree-level improved Symanzik gauge action is used.
More details of the action are described in the supplemental material of 
ref.~\cite{Colquhoun:2022atw}.
We have two different physics targets.
One is B-physics with fine lattices at zero temperature.
To this end, we have $2+1$ flavor ensembles at inverse
lattice spacings ranging from 2.45~GeV to 4.5~GeV.
The other target is finite temperature physics, to survey the phase structure
and to reveal the nature of the $U(1)_A$ symmetry across the phase
transition/cross over.  We have $2$, $2+1$, and $3$ flavors ensembles for this.
The same action is used for both zero and finite temperature targets.
We have more than 200 ensembles in total and the size is around 20~TB,
assuming we keep the configurations every 100 trajectories for the
finite temperature ensembles.

The Hybrid Monte Carlo (HMC)~\cite{Duane:1987de} algorithm is used to generate configurations.
The strange quark is treated with the rational HMC (RHMC)~\cite{Kennedy:1998cu}.
The one- or two-level Hasenbusch trick~\cite{Hasenbusch:2001ne} is combined for light quarks.
We employ the Omelyan integrator~\cite{OMELYAN2003272} for the molecular dynamics,
and apply Metropolis accept/reject test and refresh the momenta at every unit of molecular dynamics time.
We initially used the code set Iroiro++~\cite{Cossu:2013ola} and 
now we use Grid~\cite{Boyle:2015tjk} with local modifications.

Details of the up-to-date zero temperature ensembles can be found in ref.~\cite{Colquhoun:2022atw}. The lattice volume including the 5th extent is
$32^3 \times 64 \times 12$ ($1/a=2.45$~GeV) --
$64^3\times 128 \times 8$ (4.5~GeV).
Except for the finest lattice spacing, we have several quark mass points
covering a range from about 300~MeV to 500 MeV in terms of the pion mass
and one ensemble with a 230~MeV pion on a $48^3 \times 96 \times 12$ ($1/a=3.61$~GeV) lattice.
Each ensemble has 50--100 configurations, separated by 50--100 molecular dynamics trajectories.
Larger lattice volume and lighter quark mass ensembles are also planned.

Finite temperature ensembles are actively being generated and
the status at the moment of Lattice~2022 is summarized in the slides
by IK~\cite{parallel} (see also ref.~\cite{Tomiya:2016jwr,Aoki:2020noz}).
For 2-flavor ensembles, we have $N_T=8$--$14$ with the aspect ratio 2--4,
mainly for $T>T_c$. Each case has 1--6 different quark masses.
We have in total more than 50 2-flavor ensembles and most of the ensembles have 20,000 trajectories or more and are stored every 100 trajectories with some exceptions.
There are more than 70 ensembles for $2+1$ flavor finite temperature physics,
each has 20,000 trajectories and is stored every 10 trajectories (the stored configurations will be reduced).
The majority (60+) are tuned along the line of constant physics (LCP) where
the quark mass is kept fixed in physical units and the temperature is varied
in the range 120--205~MeV through changing the gauge coupling.
The quark masses are carefully tuned on some ensembles by taking into
account the effect of the residual mass of domain-wall fermions.
We currently have only one LCP parameter series with $N_T=16$  (10 ensembles)
and the rest of the LCP ensembles use $N_T=12$.
The non-LCP $2+1$ flavor ensembles employ $N_T=12$, $14$ and $16$, covering temperatures 153--204~MeV.
There are 70 ensembles for 3-flavor finite temperature physics.
The temperature is fixed to 180 MeV ($N_T=8$) or 120 MeV ($N_T=12$).
Each ensemble has 300--1200 configurations, which are stored every 10 trajectories.  We also have 3-flavor zero temperature ensembles with
the volumes $12^3 \times 24 \times 16$ and $24^3\times 48 \times 16$ for several quark masses.

The generated configurations are stored on the Japan Lattice Data Grid (JLDG)~\cite{jldg}. During the production, we also utilize the Gfarm storage system (the same as JLDG), provided by ``High Performance Computing Infrastructure (HPCI),'' which is accessible from major supercomputer sites in Japan.

The access policy is currently request based and we can provide
configurations after the relevant publications.
Every ensemble should be public in principle in the future.
Details are to be determined.

\subsection{ETMC (Extended Twisted Mass Collaboration)}
The ETM collaboration addresses mainly zero-temperature QCD physics. This includes hadron spectroscopy, hadron structure and (heavy) flavour physics. As the collaboration name suggests, the so-called Wilson clover twisted mass formulation is employed. $\mathcal{O}(a)$-improvement is realised by automatic improvement at maximal twist, while the clover term is used to further reduce the size of lattice artefacts. In the gauge sector, the Iwasaki gauge action is used.

The current simulation programme is two-fold: first, the collaboration produces gauge ensembles with $N_f=2+1+1$ dynamical quark flavours, i.e. up/down, strange and charm dynamical quarks at five values of the lattice spacing ranging from $0.091\ \mathrm{fm}$ down to $0.049\ \mathrm{fm}$. At all these lattice spacings ensembles with an (almost) physical pion mass value are available and additional ensembles with pion masses up to $350\ \mathrm{MeV}$ have been generated. The strange and charm quark masses are tuned to their physical values on all ensembles. In order to investigate finite volume effects, the collaboration has several parameter sets with only the volume varying, covering $M_\pi\cdot L$ values from $2.5$ up to $\sim 5.5$. For details. see ref.~\cite{Alexandrou:2018egz} and for a listing of the parameter choices for most ensembles, see the appendix of ref.~\cite{ExtendedTwistedMass:2021qui}.
Second, $N_f=4$ flavour ensembles are generated for each of the lattice spacing values mentioned above to be specifically used for the renormalisation effort of the collaboration. 
The overall goal of the simulation effort is to allow for a controlled continuum extrapolation at physical pion mass values. For more details, we refer to ref.~\cite{ExtendedTwistedMass:2021gbo}.

Simulations are performed using the Hybrid Monte Carlo (HMC) algorithm implemented in the tmLQCD software package~\cite{Jansen:2009xp,Deuzeman:2013xaa,Abdel-Rehim:2013wba}, which is publicly available under GPL. It implements multiple time scales and force gradient integration schemes, combined with Hasenbusch mass preconditioning for the light quarks~\cite{Urbach:2005ji}. For the strange and charm quarks a rational HMC is implemented with frequency splitting.

The DD-$\alpha$AMG~\cite{Frommer:2013fsa,Alexandrou:2016izb} multigrid iterative solver is employed for the most poorly conditioned monomials in the light sector while mixed-precision CG is used elsewhere.
In the heavy sector, multi-shift CG is used together with shift-by-shift refinement using DD-$\alpha$AMG~\cite{Alexandrou:2018wiv} for a number of smallest shifts (on machines where this is more efficient).
To target SIMD architectures, tmLQCD interfaces with the QPhiX~\cite{Joo:2013lwm} library which was extended to support the necessary operators.
Due to a more recent effort~\cite{tmLQCDGPU}, the HMC is now also able to run on GPU machines by offloading the gauge force and iterative solves using QUDA~\cite{Clark:2009wm,Babich:2011np} (including employing QUDA's MG solver~\cite{Clark:2016rdz}).
tmLQCD automatically writes gauge configurations in the ILDG format, including a header which provides some meta-data such as the creation date, the target simulation parameters, the trajectory number and the plaquette expectation value.

While the ETMC has uploaded previous sets of gauge ensembles (with $N_f=2$ and $N_f=2+1+1$ dynamical quark flavours, but without clover term) to ILDG storage elements --- see for instance refs.~\cite{Baron:2010bv,EuropeanTwistedMass:2010voq,ETM:2009ztk}, the current simulation campaign is not yet available via ILDG. However, the collaboration would immediately make use of the ILDG infrastructure as soon as a reliable service becomes available again.

The ETMC gauge ensembles are publicly available after a grace period. Thus, while all the previous gauge ensembles are available to everyone, the latest ones are currently available upon request. However, they will become public in the near future. Here, we expect to require in the order of $2.5$ PB of disc space in the foreseeable future.

\subsection{TWEXT Collaboration}
The main goal of the Twisted Wilson @ EXTreme conditions (TWEXT)
collaboration is to explore the properties of QCD at large temperatures, starting from below the chiral phase transition, up to the high temperature regime. We would like to address questions such as the QCD phase diagram, the scaling behaviour, chiral and topological properties of QCD at high temperatures and others.

We are using the Wilson twisted mass fermion discretization at maximal twist. The simulations are performed with $N_f=2+1+1$ quarks, where we keep the parameters of the strange and charm sectors close to the physical values, and we have several ensembles for pion masses starting from the physical $m_{\pi}=140$~MeV up to the heavy quark regime, $m_{\pi}=370$ MeV. For the ensembles with the physical pion mass, we are using the Wilson clover-improved twisted mass action, while in the case of higher pion masses the action is a standard Wilson twisted mass action. In our simulations we are using the tuning of the parameters by the ETM collaboration~\cite{Alexandrou:2020okk}. The simulations are carried out in a fixed scale approach, i.e., to change the system temperature we change the lattice temporal extent $N_t$, keeping the lattice spacing $a$ fixed.

Simulations are performed with the freely available tmLQCD code~\cite{tmlqcd}. In total, currently we have $60$ ensembles (one ensemble corresponds to one point in the temperature-pion~mass-lattice~spacing space), which have 50k configurations, occupying 26~TB of the disk space. 
A short summary of these ensembles can be found in ref.~\cite{Kotov:2021rah}.
We plan to add new ensembles with physical pion mass and finer lattice spacings.
Configurations are stored in the ILDG format. Ensembles are available upon request and we are open to collaboration. We plan to make TWEXT ensembles public or upload them to ILDG after some embargo time.

\subsection{PACS Collaboration}
There are two major problems in lattice QCD simulations. One is that it is still difficult to make a high precision measurement of the physical observables exclusively at the physical point. The other is that current lattice QCD simulations determine different physical observables by choosing different sets of gauge configurations. From a view point of predictability of lattice QCD it is highly desirable to make a high precision measurement of various physical observables from a unique set of gauge configurations. In order to overcome the above problems we perform very large scale simulations at the physical point towards master-field simulations named by L{\"u}scher~\cite{Luscher:2017cjh}. Lattice QCD simulations on very large lattices have inherent advantages: The statistical errors decrease thanks to the stochastic locality and the geometrical symmetries of the lattice, and the accessible minimum momentum is reduced in proportion to $1/L$ with the lattice extent $L$. We can give reliable predictions for the physical observables relevant for particle physics within and beyond the standard model.

In the past years the PACS Collaboration has been generating $2+1$ flavor QCD configurations on very large lattices at the physical point employing the stout-smeared $O(a)$-improved Wilson-clover quark action and Iwasaki gauge action. We use the stout smearing parameter $\rho=0.1$ with six smearing iterations. The improvement coefficient for the clover term is nonperturbatively determined using the Schr{\"o}dinger functional scheme. These gauge configurations, which keep the space-time volumes larger than (10 fm)$^4$, are called ``PACS10'' configurations. So far we have finished generating two gauge ensembles of (lattice spacing, lattice size)=(0.085 fm, $128^4$)~\cite{Ishikawa:2018jee} and (0.064 fm, $160^4$)~\cite{Shintani:2019wai}. We are now generating a third one of (lattice spacing, lattice size)=(0.041 fm, $256^4$) at a finer lattice spacing. The degenerate up-down quarks are simulated with the domain-decomposed HMC algorithm~\cite{Luscher:2003vf} and the strange quark with the rational HMC algorithm~\cite{Clark:2006fx}. The up-down quark determinant is separated into UV and IR parts after the even-odd preconditioning. We further apply the mass-preconditioning~\cite{Hasenbusch:2001ne} to the IR part, which is divided into three forces of $F^{\prime\prime}_{\rm IR}$, $F^{\prime}_{\rm IR}$ and ${\tilde F}_{\rm IR}$. In the end, the force terms consist of the gauge force $F_{\rm g}$, the up-down UV force $F_{\rm UV}$, the strange force $F_{\rm s}$ and the three up-down IR forces $F^{\prime\prime}_{\rm IR}$, $F^{\prime}_{\rm IR}$, ${\tilde F}_{\rm IR}$. We adopt the multiple time scale integration scheme in the molecular dynamics steps according to the hierarchical structure of $\vert\vert F_{\rm g}\vert\vert > \vert\vert F_{\rm UV}\vert\vert > \vert\vert F_{\rm s}\vert\vert \approx \vert\vert F^{\prime\prime}_{\rm IR}\vert\vert  > \vert\vert F^{\prime}_{\rm IR}\vert\vert > \vert\vert {\tilde F}_{\rm IR}\vert\vert$. The trajectory length is chosen to be $\tau\ge 1.0$. We use the mixed precision nested BiCGStab~\cite{PACS-CS:2008bkb} for the quark solver with the aid of the chronological inverter guess.  

Many gauge ensembles generated by the PACS Collaboration and the predecessors, which were the CP-PACS and PACS-CS Collaborations, are publicly available.  The current situation is available on a public webpage~\cite{jldg}. We plan to make the ``PACS10'' ensembles public through a new generation of ILDG after some embargo time. The details of the data policies are under discussion within the collaboration.

\subsection{RBC-UKQCD Collaborations}
The RIKEN-BNL Columbia (RBC) collaboration and the UKQCD collaboration,
i.e., the RBC-UKQCD collaborations, began generating $2+1$ flavor
ensembles with Domain Wall Fermions (DWF) in 2005, when the QCDOC
machines became operational at the University of Edinburgh (a 10
Tflops computer) and Brookhaven National Lab (BNL) (2 and 10 Tflops
computers).  (This followed work by the RBC collaboration using DWF
on quenched QCD configurations and preliminary investigations with
2 flavor DWF configurations that began in 1997.)  These calculations
were the first large-scale use of the recently invented Rational
Hybrid Monte Carlo algorithm by Clark and Kennedy~\cite{Clark:2003na},
which was used for the strange quark part of the evolution. Adding
this to the standard HMC algorithm for two degenerate light quarks
provided, and continues to provide, an exact evolution algorithm,
since all of the RBC-UKQCD collaborations' ensembles include the
accept/reject step at the end of each trajectory.

During the intervening years, 41 different ensembles have been
produced with either DWF or Möbius DWF (MDWF), which achieves similar
residual chiral symmetry breaking with a smaller value for $L_s$,
the extent of the fifth dimension for (M)DWF. Most of the ensembles
have been produced with the Iwasaki gauge action, which provides
for smoother gauge fields at the lattice cutoff, for a given lattice
spacing, than the Wilson action. The smoother gauge fields reduce
the residual chiral symmetry breaking for a given value of $L_s$.
The earliest ensembles have $m_\pi$ in the 200 to 400~MeV range and
volumes $\sim$ (2.5~fm)$^3$.  With improvements in algorithms and
computers, we have, over the last decade, produced 3 ensembles,
with $1/a = 1.730$, 2.359 and 2.708~GeV, $2+1$ flavors and physical
quark masses and volumes of (5.5~fm)$^3$, (5.4~fm)$^3$ and (6.9~fm)$^3$,
respectively~\cite{RBC:2014ntl}.  (The last ensemble in
this list is still being produced.)  These 3 ensembles represent
the only physical quark mass, large volume, $2+1$ flavor ensembles
generated with the chiral symmetries of the continuum essentially
intact. They support much of the wide variety of physics observables
of interest to the combined collaborations.

We also have a number of ensembles with the Iwasaki plus Dislocation
Suppressing Determinant Ratio (DSDR) gauge action, refereed to as
the ID gauge action, which suppress the copious topological tunneling
at strong coupling. The tunneling produces large residual
chiral symmetry breaking, which the ID action controls, allowing
for $2+1$ flavor simulations with physical quark masses on lattices
as coarse as $1/a = 1$~GeV.  We have ID+MDWF ensembles with
$1/a = 1$~GeV and physical volumes of (4.8~fm)$^3$, (6.4~fm)$^3$
and (9.6~fm)$^3$ --- an excellent place to investigate large
volume physics with realistic quark masses. We have also generated
ID+MDWF ensembles with G-parity boundary conditions and $1/a = 1.37$~GeV
for our ongoing work to precisely determine
Re($\epsilon'/\epsilon$)~\cite{RBC:2020kdj}.

The members of the RBC-UKQCD collaborations have interest in a broad
range of physics topics including: precision electroweak physics in light
hadrons, such as $K \to \pi \pi$ decays, $\Delta M_K$, rare kaon decays,
$K_{l3}$; heavy quark physics; the hadronic vacuum polarization
contribution to $(g-2)_\mu$ and electromagnetic contributions to
many of the above processes.

We are currently setting parameters and tuning our evolution algorithms
for $2+1+1$ flavor (M)DWF ensembles.  We are targeting a series of ensembles
with $1/a = 3$, 4 and 5~GeV and physical quark masses, using the Wilson
action. The physical volumes are about (4.5~fm)$^3$.  These will provide
a set of ensembles to allow the continuum limit to be taken with physical
quark masses and the good chiral symmetry properties of (M)DWF. This is a
long-term goal which will require the resources of the Exascale machines
that are now nearing completion.

The RBC-UKQCD collaborations make their ensembles publicly available after
an initial publication of the first physics results from these ensembles.
Earlier access can be possible for non-competing physics projects.  Many
of the ensembles are available through a Globus connection at Columbia.

\subsection{HotQCD Collaboration}
The HotQCD collaboration addresses physics questions that deal with the
properties of strong interaction matter under extreme conditions. In
particular, we are interested in universal critical behavior and the
exploration of the QCD phase diagram, the QCD equation of state at
zero and nonzero temperature, fluctuations of conserved charges and
effective degrees of freedom and the in-medium properties of hadrons,
respectively their melting. 

Due to their computational advantages and the remainder of the chiral
symmetry group, the HotQCD collaboration is predominantly using highly
improved staggered quarks (HISQ)~\cite{Follana:2006rc} with $2+1$
flavors of light and strange quarks at physical and lighter than physical
masses. The algorithm used for the generation of the gauge ensembles is the
rational hybrid Monte Carlo algorithm~\cite{Clark:2003na}. The integrator
used for the integration of the molecular dynamical trajectories has three
time scales, corresponding to the light, strange and gauge forces. For
each time scale we usually use the standard leapfrog scheme, however,
we have also implemented a 2$^{nd}$ order minimum norm integrator
(Omelyan)~\cite{OMELYAN2003272}. The acceptance is tuned to approximately
70\% on a trajectory length in the range of 0.5-1.0 molecular dynamical
time units. The code base for the calculations is now publicly available
on GitHub and was named SIMULATeQCD~\cite{SIMULATeQCD,Bollweg:2021cvl},
which stands for \textit{``a simple multi-GPU code for lattice QCD
  calculations''}. The code prints a log-file, containing the most important
provenance information. The file format of the gauge configurations is by
default NERSC, but the ILDG format is also supported. 

Recently, the HotQCD collaboration has spent the majority of
its awarded compute time on the generation of non-zero temperature
lattices of size $32^3\times 8$, $48^3\times 12$ and $64^3\times 16$,
at physical quark masses. High statistics calculations exist at 9 values
of the temperature in an interval $T\in[125,175]$ MeV~\cite{Bazavov:2020bjn,Bollweg:2021vqf,Bollweg:2022rps}. Aiming for the calculation of high order
cumulants of conserved charges, the ensemble size increased to $1.4$ mio.\
($N_\tau=8$), $~400,000$ ($N_\tau=12$) and $~20,000$ ($N_\tau=16$) gauge
configurations per temperature value. This amounts to a total of 912~TB,
938~TB and 210~TB, respectively. Currently, the gauge configurations are
stored in data projects at NERSC, ORNL, JLab and JSC. We are planning to
make these gauge configurations available through ILDG 2.0 in the
foreseeable future. In addition there exists $N_\tau=8$ ensembles at lighter
than physical quark masses, partly generated on lattices with large spatial
extent ranging up to $N_\sigma=56$ with a total size of
192~TB~\cite{HotQCD:2019xnw}.

\subsection{CLS Community Effort}
The current Coordinated Lattice Simulations (CLS) effort started in
2013 as a continuation of an earlier $N_f=2$ simulation programme
carried out by some of the member groups. CLS generate gauge ensembles with
$N_f=2+1$ flavours of non-perturbatively order $a$ improved Wilson
fermions on a tree-level Symanzik improved gauge
action. For more detail on the action,
see ref.~\cite{Bruno:2014jqa}. The lattice spacings at present range from
0.1~fm down to below 0.04~fm
with volumes from $48\cdot 24^3$ to $192\cdot 96^3$
points. The main focus is
to push lattice simulations further towards the continuum limit.
This is particularly relevant for observables involving
momenta and for heavy quark physics, however, a controlled continuum
limit is also essential for a wide range of other applications.
Ergodicity of the simulations
for lattice spacings smaller than about 0.06~fm, where
freezing of the topological charge sets in,
is achieved by employing open boundary conditions in time~\cite{Luscher:2011kk}.

All simulations are carried out using the GPL licensed package
{\sc openQCD}~\cite{Luscher:2012av,openqcd},
also in use by two of the other collaborations/initiatives who
contributed to this article.
The Hybrid Monte Carlo~\cite{Duane:1987de} (HMC) algorithm is employed and
rational HMC (RHMC)~\cite{Kennedy:1998cu}
is used for the strange quark. This is combined
with 2nd and 4th order Omelyan-Mryglod-Folk~\cite{OMELYAN2003272}
integrators for different pseudo-fermions. Moreover,
Hasenbusch frequency splitting~\cite{Hasenbusch:2001ne} and
chronological deflation acceleration of the
domain decompositioned SAP-preconditioned GCR solver are employed.
For details on the algorithmic choices and the implementation,
see refs.~\cite{Luscher:2012av,Bruno:2014jqa,Mohler:2017wnb,RQCD:2022xux}.
A small twisted mass term is added to stabilize the simulation. This,
together with the error of the rational approximation of the strange quark
determinant, is corrected for by including reweighting
factors~\cite{Luscher:2012av,Bruno:2014jqa,Mohler:2020txx} into the
expectation values. These factors
are stored, together with the gauge ensembles.

An overview of the existing ensembles can be found on a
public webpage~\cite{cls},
see also, e.g., ref.~\cite{RQCD:2022xux}. These are
redundantly stored in the {\sc openQCD} data format
at DESY Zeuthen and at Regensburg.
At present, runs exist for about 60 different combinations of the
lattice spacing, quark masses and the volume, with a total of about
130000 gauge configurations, amounting to 1~PB of data.
The log files include information on the algorithmic parameters,
the {\sc openQCD}~\cite{openqcd} version in use, the target machine,
run time and partition size,
what person was responsible for the run, as well as various control measurements
and check sums. The transfer of the configurations, log files
and other run information to the storage elements is to a large
extent handled automatically by regularly executed scripts.
These also extract detailed information that is visualized
on an internal overview/status web-page.

Many ensembles are available upon request. Further ensembles are available
upon approval by CLS, either after some embargo time or for use in
non-competing physics projects. The data policies are at present being reviewed
within CLS, with the aim to make most ensembles available through a new
generation of ILDG.

\subsection{CLQCD Collaboration}
The research interests of the CLQCD Collaboration lie in multiple areas of
lattice QCD, and we are generating three different types of gauge
configurations simultaneously.

The general-purpose ensembles are generated with the $2+1$-flavor Wilson-Clover
fermion action and the tadpole-improved Symanzik gauge action. Such
ensembles are designed to address general topics in lattice QCD like
hadron structure, the hadron spectrum and so on.
The CLQCD collaboration has generated ensembles with the tree level
tadpole improved clover action with a mild stout smearing on the gauge
link with the smearing parameter $\rho= 0.125$. For both gauge and
fermion fields, we also apply self consistent tuning for tadpole
improvement factors. Based on the above setup, we generate ensembles with
the lattice spacing ranging from about 0.1~fm to about 0.05~fm with
volumes from $24^3\times 72$ up to $48^3\times 144$ points. 

Ensembles on anisotropic lattices are generated for research on charmed
hadrons, glueballs and exotic particles. The Wilson-Clover fermion
action and the tadpole-improved Symanzik gauge action are adopted.
The anisotropy ratios of fermion and gauge fields are set to 5 to have
much finer temporal lattice spacings, while retaining a sufficiently
large spatial box size.
All the parameters are precisely tuned automatically with a set of
sophisticated QScheme scripts. One feature of these ensembles is
huge statistics; each ensemble contains 7000--10000 configurations.
We have generated $2$-flavor ensembles with $m_{\pi}^{min} = 349$~MeV,
and $2$-flavor and $2+1$-flavor ensembles with $m_{\pi} \approx 200$~MeV
are proposed.

The ensembles of $N_f=2+1$ QCD with nonzero temperature and nonzero magnetic fields are generated using the HISQ/tree action with a modified version of the SIMULATeQCD~\cite{SIMULATeQCD,Bollweg:2021cvl} code. Five different temperatures around the transition temperature and 9 different values of the magnetic field strength are realized. Additional $\sim$5k configurations of $48^3\times12$ lattices are generated for each of the 25 parameter sets. The above-mentioned configurations occupy about 120~TB.

\section{Summary}
\label{sec:summary}
\begin{table}
  \caption{Available gauge ensembles and data management.
    The numbers are only approximate. In some of the cases (e.g.,
    MILC, CLS), these only
    account for the latest
    action used by the collaboration(s)/effort/initiative.
    The storage displayed for
    JLab/W\&M/LANL/MIT/OLCF/Marseille is the
    total storage used, not only that for the gauge ensembles, while
    the storage given for ETMC includes future plans.
    \newline
    The second column indicates the public availability of the ensembles
    (0 = no, 1 = yes, after some embargo time, 2 = yes, already now)
    and the third
    column the interest in sharing these through ILDG
    (0 = no interest,  1 = interest,  2 = planned, 3 = already using).\newline
    \#ens = number of ensembles, \#cfg = total number of configurations.
    \label{tab:summary}}
  \centering
  \begin{tabular}{lccrrr}\hline
    collaboration      & public & ILDG & \#ens & \#cfg & storage (TB)\\\hline
    FASTSUM            & 1 & 1 &  25 & 22k   & 40     \\
    OpenLat            & 1 & 2 &   8 & 10k   & 30     \\
    MILC               & 1 & 1 & $>$25 & $>$50k & 650   \\
    JLab/W\&M/LANL/MIT/OLCF/Marseille & 0 & 0 &  13 & 105k   & 2000   \\
    JLQCD              & 1 & 2/3 & $>$230 & 60k & 20     \\
    ETMC               & 1 & 2/3 & 21 & 100k & 2500   \\
    TWEXT              & 1 & 1 &  60 & 50k   & 26     \\
    PACS               & 1 & 2/3 & 3 & 100   & 60     \\
    RBC-UKQCD          & 1 & 0 &  41 & 20k     & 200      \\
    HotQCD             & 1 & 2 &  58 & 15M   & 2250   \\
    CLS                & 1 & 2 &  $>$60 & 130k  & 1000   \\
    CLQCD $T=0$        & 1 & 1   & 10 & 5k   & 14     \\
    CLQCD $T>0$        & 1 & 1   & 28 & 150k & 120    \\
    HAL QCD            & 1 & 2 &  1  & 1.4k  & 70     \\
    QCDSF-UKQCD-CSSM   & 1 & 2/3 & 60 & 90k  & 300    \\\hline
  \end{tabular}
\end{table}
The fermionic actions used by the groups who responded to the call are
staggered quarks, in particular using the HISQ action (MILC, HotQCD, CLQCD),
improved Wilson quarks on isotropic
(OpenLat, JLab/WM/LANL/MIT/OLCF/Marseille,
PACS, CLS, CLQCD, HAL QCD, QCDSF-UKQCD-CSSM)
and anisotropic lattices (FASTSUM, CLQCD),
twisted mass fermions (ETMC, TWEXT) and Domain Wall fermions
(JLQCD, RBC-UKQCD). Some collaborations have generated ensembles
exclusively with temperatures $T>0$ (TWEXT, HotQCD, CLQCD),
others both for $T>0$ and for $T=0$ (FASTSUM, MILC) and the
remaining ones only for $T=0$.
The total number of gauge ensembles/configurations and
the storage requirements, public availability and ILDG plans
are summarized in Table~\ref{tab:summary}.
For details on the action, simulation setup, data management strategy etc.,
we refer to the individual contributions in the previous section.
After the session was filled, additional input
was received from Takumi Doi on behalf of the HAL~QCD Collaboration
and from James Zanotti on behalf of QCDSF-UKQCD-CSSM. This
is also included in the table.

The collected material is a snapshot of the
activities of a large fraction of the ``data providers'' generating
$N_f=2+1$ and $N_f=2+1+1$ QCD ensembles.
We hope that this information on ongoing ensemble generation campaigns,
the associated usage and access policies as well as the storage requirements
is of use, both to ``data providers'' and to ``data consumers.''
The intent is that this summary
encourages communication, coordination and best
practice within the community and beyond.

Most participants of the parallel session indicated an interest in
sharing gauge configurations through ILDG, see the third
column of Table~\ref{tab:summary}.
This may encourage and assist the efforts to modernize and extend
ILDG as discussed at the ILDG~lunch~\cite{parallel} and in a
plenary contribution~\cite{Karsch:2022tqw} at this conference.
In helping to establish well-defined workflows and minimum
quality standards for data management, ILDG can provide an
important framework to support data sharing.
However, acquiring storage space, where data can be held accessible
(according to the envisaged access policies)
and persistent for more than a decade, remains the responsibility of
the regional grids and the
contributing collaborations (irrespective of whether they join the ILDG
effort or not).
Many different regional funding agencies are involved.
Therefore, this challenge --- as well as supporting the necessary scientific
personnel --- cannot be addressed globally with one and the same
strategy.
However, local experiences and information can be shared and
activities of the community as a whole can be coordinated.

\acknowledgments
Gunnar Bali thanks Hubert Simma for discussions and help in the organization
of the session. For lack of space, we only acknowledge funding received by the
co-authors and not by all the members of the respective collaborations. We also
refrain from a detailed listing of the respective
computer time grants, which are acknowledged in the publications cited
in the text. Instead, we list the high performance computers that
were used to generate the gauge ensembles. Our thanks go to
all the institutions that run these machines, their user support
and the funding agencies.

Gunnar Bali, Christian Schmidt and Wolfgang Söldner acknowledge support by the Deutsche Forschungsgemeinschaft (DFG) under the grant
``NFDI 39/1'' (PUNCH4NFDI). Gunnar Bali and Wolfgang Söldner acknowledge support from the European
Union’s Horizon 2020
research and innovation programme under grant agreements~813942 (ITN EuroPLEx)
and~824093 (STRONG-2020).
Ryan Bignell is supported by Science and Technology Facilities Council (STFC) grant ST/000813/1.
Anthony Francis is supported by the Japanese Ministry of Science and Technology Taiwan (MOST) under grant 111\nobreakdash-2112\nobreakdash-M\nobreakdash-A49\nobreakdash-018\nobreakdash-MY2.
Steven Gottlieb is supported by the U.S.\ Department of Energy (DOE) under
Award No.~DE\nobreakdash-SC0010120.
The work of Rajan Gupta is supported by the DOE High Energy Physics program (DOE~HEP) under Award No.~DE\nobreakdash-AC52\nobreakdash-06NA25396 and by the LANL~LDRD program.
Issaku Kanamori is partially supported by the Ministry of Education, Culture, Sports, Science and Technology (MEXT) in the ``Program for Promoting Researches on the Supercomputer Fugaku'' (Simulation for basic science: from fundamental laws of particles to creation of nuclei). He also acknowledges the Japanese Society for the Promotion of Science (JSPS) KAKENHI grant (JP20K03961).
The ``PACS10'' configuration generation by the PACS Collaboration (presented
here by Yoshinobu Kuramashi) is
supported in part by Grants-in-Aid for Scientific Research from the MEXT
(Grant No.~19H01892).
Robert Mawhinney is supported in part by the DOE grant DE\nobreakdash-SC0011941.
Christian Schmidt is supported through the DFG Collaborative Research Centre
SFB/TRR~211 as well as by the European Union grant 283286.

The participating collaborations have generated their gauge ensembles
(list sorted alphabetically according to country, institute, computer)\\
on the HPC Cluster of the Institute of Theoretical Physics of the Chinese Academy of Sciences (CAS) in Beijing,\\
on computers at the Nuclear Science Computing Center (NSC3) at Central China Normal University (CCNU) in Wuhan,\\
on computers at the Southern Nuclear Science Computing Center (SNSC) at South China Normal University (SCNU) in Guangzhou,\\
on Siyuan Mark 1 at the Center for High Performance Computing at Shanghai Jiao Tong University (SJTU), China,\\
on Occigen at Centre Informatique National de l’Enseignement Supérieur (C.I.N.E.S.) in Montpellier,\\
on Jean Zay at Institut du Développement et des Ressources en Informatique
Scientifique (IDRIS) in Orsay,\\
on Ir\`ene Joliot-Curie at Très Grand Centre de Calcul (TGCC) in Bruyères-le-Châtel, France,\\
on Goethe-HLR and Loewe-CSC at Goethe-Universität Frankfurt,\\
on Clover and HIMster-II at Helmholtz Institute Mainz,\\
on Hazel~Hen and HAWK at Höchstleistungsrechenzentrum Stuttgart (HLRS),\\
on Mogon-II at Johannes-Gutenberg-Universität Mainz,\\
on JUGEEN, JUQUEEN, JURECA, JURECA-Booster, JUWELS and JUWELS-Booster at Jülich Supercomputing Centre (JSC),\\
on SuperMUC and SuperMUC-NG at Leibniz Rechenzentrum (LRZ) in Garching,\\
on Oculus, Nocutua and Nocutua2 at Paderborn Center for Parallel Computing (PC2),\\
on Fritz at Regionales Rechenzentrum Erlangen (RRZE),\\
on the Bielefeld GPU-Cluster of Universität Bielefeld,\\
on Bonna at Universität Bonn,\\
on Lise (HLRN-IV) at Zuse-Institut Berlin (ZIB),\\
on iDataCool and Athene~2 at Universität Regensburg, Germany,\\
on Stokes at the Irish Centre for High-End Computing (ICHEC) in Galway, Ireland,\\
on Fermi, Marconi~A1, Marconi~A2, Marconi~A3 and Marconi~100 at CINECA in Bologna, Italy,\\
on Polaire and Grand Chariot at Hokkaido University in Sapporo,\\
on IBM Blue Gene/Q (BG/Q) computers at the High Energy Accelerator Research Organization (KEK) in Tsukuba,\\
on SQUID at Osaka University,\\
on Supercomputer Fugaku at the RIKEN Center for Computational Science (R-CCS) in Kobe,\\
on Oakforest-PACS at the Joint Center for Advanced High Performance Computing (JCAHPC) of the Universities of Tokyo and Tsukuba in Kashiwa,\\
on Wisteria at the Information Technology Center of the University of Tokyo in Kashiwa, Japan,\\
on Prometheus at Akademickie Centrum Komputerowe CYFRONET in Kraków,\\
on Okeanos at ICM Uniwersytet Warszawski, Poland,\\
on hpc-qcd at CERN in Geneva,\\
on Piz~Daint at the Centro Svizzero di Calcolo Scientifico (CSCS) in Lugano, Switzerland,\\
on Tesseract and Tursa of the DiRAC Extreme Scaling service in Edinburgh,\\
on the BG/L, BG/Q and QCDOC at the University of Edinburgh,\\
on Sunbird of Supercomputing Wales at Swansea University, UK,\\
on Intrepid, Mira and Theta of the Argonne Leadership Computing Facility (ALCF) at the Argonne National Laboratory (ANL) in Lemont,\\
on the BG/Q, QCDOC and QCDSP of the RIKEN-BNL Research Center, the BG/L of the New York Center for Computational Sciences (NYCCS) and the QCDOC of USQCD at Brookhaven National Laboratory (BNL) in Upton,\\
on the QCDSP at Columbia University in New York,\\
on the GPU and KNL Clusters of USQCD at Jefferson Lab (JLab) in Newport News and at the Fermi National Accelerator Laboratory (FNAL) in Batavia,\\
on Big Red 2+, Big Red 3, and Big Red 200 at Indiana University in Bloomington,\\
on Sequoia and Vulcan at Lawrence Livermore National Lab (LLNL) in Livermore,\\
on Badger, Chicoma and Grizzly at Los Alamos National Laboratory (LANL) Institutional Computing,\\
on Blue Waters of the National Center for Supercomputing Applications (NCSA) at the University of Illinois in Urbana-Champaign,\\
on Cori, Edison and Perlmutter at the National Energy Research Scientific Computing Center (NERSC) in Berkeley,\\
on computers of the National Center for Atmospheric Research (NCAR) in Boulder,\\
on Summit and Titan at Oak Ridge Leadership Class Facility (OLCF)
at Oak Ridge National Laboratory (ORNL),\\
on computers of the National Institute for Computational Sciences (NICS) of ORNL and the University of Tennessee at ORNL,\\
on Frontera and Stampede~2 at Texas Advanced Computing Center (TACC) in Austin\\
and on RMACC Summit at the University of Colorado in Boulder, USA.
\setlength{\bibsep}{0pt plus 0.3ex}
\providecommand{\href}[2]{#2}\begingroup\raggedright\endgroup

\begin{thebibliography}{10}
\bibitem{HadronSpectrum:2009krc}
{\scshape Hadron Spectrum} collaboration, M.~Peardon et~al., \emph{{A Novel
  quark-field creation operator construction for hadronic physics in lattice
  QCD}}, \href{https://doi.org/10.1103/PhysRevD.80.054506}{\emph{Phys. Rev. D}
  {\bfseries 80} (2009) 054506}
  [\href{https://arxiv.org/abs/0905.2160}{{\ttfamily 0905.2160}}].

\bibitem{Davies:2002mu}
{\scshape UKQCD} collaboration, C.~T.~H. Davies, A.~C. Irving, R.~D. Kenway and
  C.~M. Maynard, \emph{{International Lattice Data Grid}},
  \href{https://doi.org/10.1016/S0920-5632(03)01509-3}{\emph{Nucl. Phys. B
  Proc. Suppl.} {\bfseries 119} (2003) 225}
  [\href{https://arxiv.org/abs/hep-lat/0209121}{{\ttfamily hep-lat/0209121}}].

\bibitem{Irving:2008uk}
T.~Yoshie, \emph{{Making use of the International Lattice Data Grid}},
  \href{https://doi.org/10.22323/1.066.0019}{\emph{PoS} {\bfseries LATTICE2008}
  (2008) 019} [\href{https://arxiv.org/abs/0812.0849}{{\ttfamily 0812.0849}}].

\bibitem{Maynard:2009szr}
C.~M. Maynard, \emph{{International Lattice Data Grid: Turn on, plug in and
  download}}, \href{https://doi.org/10.22323/1.091.0020}{\emph{PoS} {\bfseries
  LAT2009} (2009) 020} [\href{https://arxiv.org/abs/1001.5207}{{\ttfamily
  1001.5207}}].

\bibitem{Beckett:2009cb}
M.~G. Beckett, B.~Joo, C.~M. Maynard, D.~Pleiter, O.~Tatebe and T.~Yoshie,
  \emph{{Building the International Lattice Data Grid}},
  \href{https://doi.org/10.1016/j.cpc.2011.01.027}{\emph{Comput. Phys. Commun.}
  {\bfseries 182} (2011) 1208}
  [\href{https://arxiv.org/abs/0910.1692}{{\ttfamily 0910.1692}}].

\bibitem{Wilkinson2016}
M.~Wilkinson, M.~Dumontier, I.~Aalbersberg et~al., \emph{{The FAIR Guiding
  Principles for scientific data management and stewardship}},
  \href{https://doi.org/10.1038/sdata.2016.18}{\emph{Sci Data} {\bfseries 3}
  (2016) 160018}.

\bibitem{Coddington:2007gz}
{\scshape ILDG Metadata Working Group} collaboration, P.~Coddington, B.~Joo,
  C.~M. Maynard, D.~Pleiter and T.~Yoshie, \emph{{Marking up lattice QCD
  configurations and ensembles}},
  \href{https://doi.org/10.22323/1.042.0048}{\emph{PoS} {\bfseries LATTICE2007}
  (2007) 048} [\href{https://arxiv.org/abs/0710.0230}{{\ttfamily 0710.0230}}].

\bibitem{Karsch:2022tqw}
F.~Karsch, H.~Simma and T.~Yoshie, \emph{{The International Lattice Data Grid
  -- towards FAIR Data}},  \href{https://arxiv.org/abs/2212.08392}{{\ttfamily
  2212.08392}}.

\bibitem{parallel}
S.~Gottlieb et~al., ``Parallel session at {L}attice~2021: {L}attice {D}ata.''
  \url{https://indico.hiskp.uni-bonn.de/event/40/sessions/98/#20220809}.

\bibitem{Edwards:2008ja}
R.~G. Edwards, B.~Joó and H.-W. Lin, \emph{{Tuning for Three-flavors of
  Anisotropic Clover Fermions with Stout-link Smearing}},
  \href{https://doi.org/10.1103/PhysRevD.78.054501}{\emph{Phys. Rev. D}
  {\bfseries 78} (2008) 054501}
  [\href{https://arxiv.org/abs/0803.3960}{{\ttfamily 0803.3960}}].

\bibitem{HadronSpectrum:2008xlg}
{\scshape Hadron Spectrum} collaboration, H.-W. Lin et~al., \emph{{First
  results from $2+1$ dynamical quark flavors on an anisotropic lattice:
  Light-hadron spectroscopy and setting the strange-quark mass}},
  \href{https://doi.org/10.1103/PhysRevD.79.034502}{\emph{Phys. Rev. D}
  {\bfseries 79} (2009) 034502}
  [\href{https://arxiv.org/abs/0810.3588}{{\ttfamily 0810.3588}}].

\bibitem{Edwards:2004sx}
{\scshape SciDAC, LHPC, UKQCD} collaboration, R.~G. Edwards and B.~Joó,
  \emph{{The Chroma software system for lattice QCD}},
  \href{https://doi.org/10.1016/j.nuclphysbps.2004.11.254}{\emph{Nucl. Phys.
  Proc. Suppl.} {\bfseries 140} (2005) 832}
  [\href{https://arxiv.org/abs/hep-lat/0409003}{{\ttfamily hep-lat/0409003}}].

\bibitem{glesaaen_jonas_rylund_2018_2216355}
J.~R. Glesaaen and B.~Jäger, ``openqcd-fastsum.''
  \url{https://gitlab.com/fastsum}, Apr., 2018.
\newblock 10.5281/zenodo.2216355.

\bibitem{Luscher:2012av}
M.~Lüscher and S.~Schaefer, \emph{{Lattice QCD with open boundary conditions
  and twisted-mass reweighting}},
  \href{https://doi.org/10.1016/j.cpc.2012.10.003}{\emph{Comput. Phys. Commun.}
  {\bfseries 184} (2013) 519}
  [\href{https://arxiv.org/abs/1206.2809}{{\ttfamily 1206.2809}}].

\bibitem{openqcd}
``{\texttt{openQCD}: Simulation programs for lattice QCD}.''
  \url{https://luscher.web.cern.ch/luscher/openQCD/}.

\bibitem{Aarts:2014nba}
G.~Aarts, C.~Allton, A.~Amato, P.~Giudice, S.~Hands and J.-I. Skullerud,
  \emph{{Electrical conductivity and charge diffusion in thermal QCD from the
  lattice}}, \href{https://doi.org/10.1007/JHEP02(2015)186}{\emph{JHEP}
  {\bfseries 02} (2015) 186} [\href{https://arxiv.org/abs/1412.6411}{{\ttfamily
  1412.6411}}].

\bibitem{Aarts:2020vyb}
G.~Aarts et~al., \emph{{Properties of the QCD thermal transition with Nf=2+1
  flavors of Wilson quark}},
  \href{https://doi.org/10.1103/PhysRevD.105.034504}{\emph{Phys. Rev. D}
  {\bfseries 105} (2022) 034504}
  [\href{https://arxiv.org/abs/2007.04188}{{\ttfamily 2007.04188}}].

\bibitem{zenodo}
{European Organization for Nuclear Research} and {OpenAIRE}, ``Zenodo.''
  \url{https://www.zenodo.org/}.

\bibitem{Francis:2019muy}
A.~Francis, P.~Fritzsch, M.~Lüscher and A.~Rago, \emph{{Master-field
  simulations of O($a$)-improved lattice QCD: Algorithms, stability and
  exactness}}, \href{https://doi.org/10.1016/j.cpc.2020.107355}{\emph{Comput.
  Phys. Commun.} {\bfseries 255} (2020) 107355}
  [\href{https://arxiv.org/abs/1911.04533}{{\ttfamily 1911.04533}}].

\bibitem{Francis:2022hyr}
A.~S. Francis, F.~Cuteri, P.~Fritzsch, G.~Pederiva, A.~Rago, A.~Shindler,
  A.~Walker-Loud and S.~Zafeiropoulos, \emph{{Properties, ensembles and hadron
  spectra with Stabilised Wilson Fermions}},
  \href{https://doi.org/10.22323/1.396.0118}{\emph{PoS} {\bfseries LATTICE2021}
  (2022) 118} [\href{https://arxiv.org/abs/2201.03874}{{\ttfamily
  2201.03874}}].

\bibitem{Luscher:2011kk}
M.~Lüscher and S.~Schaefer, \emph{{Lattice QCD without topology barriers}},
  \href{https://doi.org/10.1007/JHEP07(2011)036}{\emph{JHEP} {\bfseries 07}
  (2011) 036} [\href{https://arxiv.org/abs/1105.4749}{{\ttfamily 1105.4749}}].

\bibitem{Luscher:2017cjh}
M.~Lüscher, \emph{{Stochastic locality and master-field simulations of very
  large lattices}},
  \href{https://doi.org/10.1051/epjconf/201817501002}{\emph{EPJ Web Conf.}
  {\bfseries 175} (2018) 01002}
  [\href{https://arxiv.org/abs/1707.09758}{{\ttfamily 1707.09758}}].

\bibitem{Luscher:2010iy}
M.~Lüscher, \emph{{Properties and uses of the Wilson flow in lattice QCD}},
  \href{https://doi.org/10.1007/JHEP08(2010)071}{\emph{JHEP} {\bfseries 08}
  (2010) 071} [\href{https://arxiv.org/abs/1006.4518}{{\ttfamily 1006.4518}}],
  [Erratum: JHEP 03, 092 (2014)].

\bibitem{Bruno:2016plf}
M.~Bruno, T.~Korzec and S.~Schaefer, \emph{{Setting the scale for the CLS $2 +
  1$ flavor ensembles}},
  \href{https://doi.org/10.1103/PhysRevD.95.074504}{\emph{Phys. Rev. D}
  {\bfseries 95} (2017) 074504}
  [\href{https://arxiv.org/abs/1608.08900}{{\ttfamily 1608.08900}}].

\bibitem{Cuteri:2022gmg}
F.~Cuteri, A.~Francis, P.~Fritzsch, G.~Pederiva, A.~Rago, A.~Shindler,
  A.~Walker-Loud and S.~Zafeiropoulos, \emph{{Gauge generation and
  dissemination in OpenLat}},
  \href{https://arxiv.org/abs/2212.07314}{{\ttfamily 2212.07314}}.

\bibitem{openlat}
``{Openlat online repository and point of reference}.''
  \url{https://openlat1.gitlab.io}.

\bibitem{Follana:2006rc}
{\scshape HPQCD, UKQCD} collaboration, E.~Follana et~al., \emph{{Highly
  improved staggered quarks on the lattice, with applications to charm
  physics}}, \href{https://doi.org/10.1103/PhysRevD.75.054502}{\emph{Phys. Rev.
  D} {\bfseries 75} (2007) 054502}
  [\href{https://arxiv.org/abs/hep-lat/0610092}{{\ttfamily hep-lat/0610092}}].

\bibitem{MILC:2009mpl}
{\scshape MILC} collaboration, A.~Bazavov et~al., \emph{{Nonperturbative QCD
  simulations with $2+1$ flavors of improved staggered quarks}},
  \href{https://doi.org/10.1103/RevModPhys.82.1349}{\emph{Rev. Mod. Phys.}
  {\bfseries 82} (2010) 1349}
  [\href{https://arxiv.org/abs/0903.3598}{{\ttfamily 0903.3598}}].

\bibitem{Miller:2020evg}
N.~Miller et~al., \emph{{Scale setting the M\"obius domain wall fermion on
  gradient-flowed HISQ action using the omega baryon mass and the gradient-flow
  scales $t_0$ and $w_0$}},
  \href{https://doi.org/10.1103/PhysRevD.103.054511}{\emph{Phys. Rev. D}
  {\bfseries 103} (2021) 054511}
  [\href{https://arxiv.org/abs/2011.12166}{{\ttfamily 2011.12166}}].

\bibitem{Clark:2003na}
M.~A. Clark and A.~D. Kennedy, \emph{{The RHMC algorithm for two flavors of
  dynamical staggered fermions}},
  \href{https://doi.org/10.1016/S0920-5632(03)02732-4}{\emph{Nucl. Phys. B
  Proc. Suppl.} {\bfseries 129} (2004) 850}
  [\href{https://arxiv.org/abs/hep-lat/0309084}{{\ttfamily hep-lat/0309084}}].

\bibitem{Hasenbusch:2001ne}
M.~Hasenbusch, \emph{{Speeding up the hybrid Monte Carlo algorithm for
  dynamical fermions}},
  \href{https://doi.org/10.1016/S0370-2693(01)01102-9}{\emph{Phys. Lett. B}
  {\bfseries 519} (2001) 177}
  [\href{https://arxiv.org/abs/hep-lat/0107019}{{\ttfamily hep-lat/0107019}}].

\bibitem{MILC:2010pul}
{\scshape MILC} collaboration, A.~Bazavov et~al., \emph{{Scaling studies of QCD
  with the dynamical HISQ action}},
  \href{https://doi.org/10.1103/PhysRevD.82.074501}{\emph{Phys. Rev. D}
  {\bfseries 82} (2010) 074501}
  [\href{https://arxiv.org/abs/1004.0342}{{\ttfamily 1004.0342}}].

\bibitem{MILC:2012znn}
{\scshape MILC} collaboration, A.~Bazavov et~al., \emph{{Lattice QCD Ensembles
  with four flavors of highly improved staggered quarks}},
  \href{https://doi.org/10.1103/PhysRevD.87.054505}{\emph{Phys. Rev. D}
  {\bfseries 87} (2013) 054505}
  [\href{https://arxiv.org/abs/1212.4768}{{\ttfamily 1212.4768}}].

\bibitem{USQCD_SciDAC}
``{\texttt{USQCD}} software page.'' \url{https://www.usqcd.org/software.html}.

\bibitem{Joo:2013lwm}
B.~Joó, D.~D. Kalamkar, K.~Vaidyanathan, M.~Smelyanskiy, K.~Pamnany, V.~W.
  Lee, P.~Dubey and W.~Watson, \emph{{Lattice QCD on Intel\textregistered{}
  Xeon Phi coprocessors}},
  \href{https://doi.org/10.1007/978-3-642-38750-0_4}{\emph{Lect. Notes Comput.
  Sci.} {\bfseries 7905} (2013) 40}.

\bibitem{Li:2014kxa}
R.~Li and S.~Gottlieb, \emph{{Staggered Dslash performance on Intel Xeon Phi
  architecture}}, \href{https://doi.org/10.22323/1.214.0034}{\emph{PoS}
  {\bfseries LATTICE2014} (2015) 034}
  [\href{https://arxiv.org/abs/1411.2087}{{\ttfamily 1411.2087}}].

\bibitem{DeTar:2016ndn}
C.~DeTar, D.~Doerfler, S.~Gottlieb, A.~Jha, D.~Kalamkar, R.~Li and
  D.~Toussaint, \emph{{MILC staggered conjugate gradient performance on Intel
  KNL}}, \href{https://doi.org/10.22323/1.256.0270}{\emph{PoS} {\bfseries
  LATTICE2016} (2016) 270} [\href{https://arxiv.org/abs/1611.00728}{{\ttfamily
  1611.00728}}].

\bibitem{DeTar:2018pyj}
C.~DeTar, S.~Gottlieb, R.~Li and D.~Toussaint, \emph{{MILC code performance on
  high end CPU and GPU supercomputer clusters}},
  \href{https://doi.org/10.1051/epjconf/201817502009}{\emph{EPJ Web Conf.}
  {\bfseries 175} (2018) 02009}.

\bibitem{Barros:2008rd}
K.~Barros, R.~Babich, R.~Brower, M.~A. Clark and C.~Rebbi, \emph{{Blasting
  through lattice calculations using CUDA}},
  \href{https://doi.org/10.22323/1.066.0045}{\emph{PoS} {\bfseries LATTICE2008}
  (2008) 045} [\href{https://arxiv.org/abs/0810.5365}{{\ttfamily 0810.5365}}].

\bibitem{Clark:2009wm}
M.~A. Clark, R.~Babich, K.~Barros, R.~C. Brower and C.~Rebbi, \emph{{Solving
  Lattice QCD systems of equations using mixed precision solvers on GPUs}},
  \href{https://doi.org/10.1016/j.cpc.2010.05.002}{\emph{Comput. Phys. Commun.}
  {\bfseries 181} (2010) 1517}
  [\href{https://arxiv.org/abs/0911.3191}{{\ttfamily 0911.3191}}].

\bibitem{Babich:2011np}
R.~Babich, M.~A. Clark, B.~Joo, G.~Shi, R.~C. Brower and S.~Gottlieb,
  \emph{{Scaling Lattice QCD beyond 100 GPUs}},  in \emph{{SC11: International
  Conference for High Performance Computing, Networking, Storage and
  Analysis}}, 9, 2011, \href{https://arxiv.org/abs/1109.2935}{{\ttfamily
  1109.2935}}, \href{https://doi.org/10.1145/2063384.2063478}{DOI}.

\bibitem{QUDADownload}
M.~A. Clark and R.~Babich, ``{QUDA: A library for QCD on GPUs}.''
  \url{http://lattice.github.io/quda/}.

\bibitem{MILCwiki}
``{\texttt{MILC}} wiki page.''
  \url{https://github.com/milc-qcd/sharing/wiki/LatticeSharing }.

\bibitem{Clark:2016rdz}
M.~A. Clark, B.~Joó, A.~Strelchenko, M.~Cheng, A.~Gambhir and R.~Brower,
  \emph{{Accelerating lattice QCD multigrid on GPUs using fine-grained
  parallelization}},  in \emph{SC16: Proceedings of the International
  Conference for High Performance Computing, Networking, Storage and Analysis},
  p.~795, 12, 2016, \href{https://arxiv.org/abs/1612.07873}{{\ttfamily
  1612.07873}}, \href{https://doi.org/10.1109/SC.2016.67}{DOI}.

\bibitem{ChromaDownload}
R.~G. Edwards and B.~Joó, ``{The Chroma Software System for LatticeQCD}.''
  \url{http://github.com/jeffersonlab/chroma}.

\bibitem{Winter:2014dka}
F.~T. Winter, M.~A. Clark, R.~G. Edwards and B.~Jo\'o, \emph{{A framework for
  lattice QCD calculations on GPUs}},  in \emph{{28th IEEE International
  Parallel and Distributed Processing Symposium}}, 8, 2014,
  \href{https://arxiv.org/abs/1408.5925}{{\ttfamily 1408.5925}},
  \href{https://doi.org/10.1109/IPDPS.2014.112}{DOI}.

\bibitem{QDPJITDownload}
F.~Winter, ``{QDP-JIT Download}.''
  \url{http://github.com/jeffersonlab/qdp-jit}.

\bibitem{Brower:2012vk}
R.~C. Brower, H.~Neff and K.~Orginos, \emph{{The M\"obius domain wall fermion
  algorithm}}, \href{https://doi.org/10.1016/j.cpc.2017.01.024}{\emph{Comput.
  Phys. Commun.} {\bfseries 220} (2017) 1}
  [\href{https://arxiv.org/abs/1206.5214}{{\ttfamily 1206.5214}}].

\bibitem{Colquhoun:2022atw}
{\scshape JLQCD} collaboration, B.~Colquhoun, S.~Hashimoto, T.~Kaneko and
  J.~Koponen, \emph{{Form factors of $B\rightarrow\pi\ell\nu$ and a
  determination of $|V_{ub}|$ with M\"obius domain-wall fermions}},
  \href{https://doi.org/10.1103/PhysRevD.106.054502}{\emph{Phys. Rev. D}
  {\bfseries 106} (2022) 054502}
  [\href{https://arxiv.org/abs/2203.04938}{{\ttfamily 2203.04938}}].

\bibitem{Duane:1987de}
S.~Duane, A.~D. Kennedy, B.~J. Pendleton and D.~Roweth, \emph{{Hybrid Monte
  Carlo}}, \href{https://doi.org/10.1016/0370-2693(87)91197-X}{\emph{Phys.
  Lett. B} {\bfseries 195} (1987) 216}.

\bibitem{Kennedy:1998cu}
A.~D. Kennedy, I.~Horvath and S.~Sint, \emph{{A new exact method for dynamical
  fermion computations with nonlocal actions}},
  \href{https://doi.org/10.1016/S0920-5632(99)85217-7}{\emph{Nucl. Phys. B
  Proc. Suppl.} {\bfseries 73} (1999) 834}
  [\href{https://arxiv.org/abs/hep-lat/9809092}{{\ttfamily hep-lat/9809092}}].

\bibitem{OMELYAN2003272}
I.~Omelyan, I.~Mryglod and R.~Folk, \emph{Symplectic analytically integrable
  decomposition algorithms: classification, derivation, and application to
  molecular dynamics, quantum and celestial mechanics simulations},
  \href{https://doi.org/https://doi.org/10.1016/S0010-4655(02)00754-3}{\emph{Comput.
  Phys. Commun.} {\bfseries 151} (2003) 272}.

\bibitem{Cossu:2013ola}
G.~Cossu, J.~Noaki, S.~Hashimoto, T.~Kaneko, H.~Fukaya, P.~A. Boyle and J.~Doi,
  \emph{{JLQCD IroIro++ lattice code on BG/Q}},
  \href{https://doi.org/10.22323/1.187.0482}{\emph{PoS} {\bfseries Lattice
  2013} (2014) } [\href{https://arxiv.org/abs/1311.0084}{{\ttfamily
  1311.0084}}].

\bibitem{Boyle:2015tjk}
P.~Boyle, A.~Yamaguchi, G.~Cossu and A.~Portelli, \emph{{Grid: A next
  generation data parallel C++ QCD library}},
  \href{https://arxiv.org/abs/1512.03487}{{\ttfamily 1512.03487}}.

\bibitem{Tomiya:2016jwr}
{\scshape JLQCD} collaboration, A.~Tomiya, G.~Cossu, S.~Aoki, .~Fukaya,
  S.~Hashimoto, T.~Kaneko and J.~Noaki, \emph{{Evidence of effective axial
  $U(1)$ symmetry restoration at high temperature QCD}},
  \href{https://doi.org/10.1103/PhysRevD.96.034509}{\emph{Phys. Rev. D}
  {\bfseries 96} (2017) 034509}
  [\href{https://arxiv.org/abs/1612.01908}{{\ttfamily 1612.01908}}], [Addendum:
  Phys.Rev.D 96, 079902 (2017)].

\bibitem{Aoki:2020noz}
{\scshape JLQCD} collaboration, S.~Aoki, Y.~Aoki, G.~Cossu, H.~Fukaya,
  S.~Hashimoto, T.~Kaneko, C.~Rohrhofer and K.~Suzuki, \emph{{Study of the
  axial $U(1)$ anomaly at high temperature with lattice chiral fermions}},
  \href{https://doi.org/10.1103/PhysRevD.103.074506}{\emph{Phys. Rev. D}
  {\bfseries 103} (2021) 074506}
  [\href{https://arxiv.org/abs/2011.01499}{{\ttfamily 2011.01499}}].

\bibitem{jldg}
``{Japan Lattice Data Grid}.'' \url{https://www.jldg.org/}.

\bibitem{Alexandrou:2018egz}
C.~Alexandrou et~al., \emph{{Simulating twisted mass fermions at physical
  light, strange and charm quark masses}},
  \href{https://doi.org/10.1103/PhysRevD.98.054518}{\emph{Phys. Rev. D}
  {\bfseries 98} (2018) 054518}
  [\href{https://arxiv.org/abs/1807.00495}{{\ttfamily 1807.00495}}].

\bibitem{ExtendedTwistedMass:2021qui}
{\scshape ETM} collaboration, C.~Alexandrou et~al., \emph{{Ratio of kaon and
  pion leptonic decay constants with $N_f=2+1+1$ Wilson-clover twisted-mass
  fermions}}, \href{https://doi.org/10.1103/PhysRevD.104.074520}{\emph{Phys.
  Rev. D} {\bfseries 104} (2021) 074520}
  [\href{https://arxiv.org/abs/2104.06747}{{\ttfamily 2104.06747}}].

\bibitem{ExtendedTwistedMass:2021gbo}
{\scshape ETM} collaboration, C.~Alexandrou et~al., \emph{{Quark masses using
  twisted-mass fermion gauge ensembles}},
  \href{https://doi.org/10.1103/PhysRevD.104.074515}{\emph{Phys. Rev. D}
  {\bfseries 104} (2021) 074515}
  [\href{https://arxiv.org/abs/2104.13408}{{\ttfamily 2104.13408}}].

\bibitem{Jansen:2009xp}
K.~Jansen and C.~Urbach, \emph{{tmLQCD: A Program suite to simulate Wilson
  twisted mass lattice QCD}},
  \href{https://doi.org/10.1016/j.cpc.2009.05.016}{\emph{Comput. Phys. Commun.}
  {\bfseries 180} (2009) 2717}
  [\href{https://arxiv.org/abs/0905.3331}{{\ttfamily 0905.3331}}].

\bibitem{Deuzeman:2013xaa}
A.~Deuzeman, K.~Jansen, B.~Kostrzewa and C.~Urbach, \emph{{Experiences with
  OpenMP in tmLQCD}}, \href{https://doi.org/10.22323/1.187.0416}{\emph{PoS}
  {\bfseries LATTICE2013} (2014) 416}
  [\href{https://arxiv.org/abs/1311.4521}{{\ttfamily 1311.4521}}].

\bibitem{Abdel-Rehim:2013wba}
A.~Abdel-Rehim, F.~Burger, A.~Deuzeman, K.~Jansen, B.~Kostrzewa, L.~Scorzato
  and C.~Urbach, \emph{{Recent developments in the tmLQCD software suite}},
  \href{https://doi.org/10.22323/1.187.0414}{\emph{PoS} {\bfseries LATTICE2013}
  (2014) 414} [\href{https://arxiv.org/abs/1311.5495}{{\ttfamily 1311.5495}}].

\bibitem{Urbach:2005ji}
C.~Urbach, K.~Jansen, A.~Shindler and U.~Wenger, \emph{{HMC algorithm with
  multiple time scale integration and mass preconditioning}},
  \href{https://doi.org/10.1016/j.cpc.2005.08.006}{\emph{Comput. Phys. Commun.}
  {\bfseries 174} (2006) 87}
  [\href{https://arxiv.org/abs/hep-lat/0506011}{{\ttfamily hep-lat/0506011}}].

\bibitem{Frommer:2013fsa}
A.~Frommer, K.~Kahl, S.~Krieg, B.~Leder and M.~Rottmann, \emph{{Adaptive
  aggregation based domain decomposition multigrid for the lattice Wilson Dirac
  operator}}, \href{https://doi.org/10.1137/130919507}{\emph{SIAM J. Sci.
  Comput.} {\bfseries 36} (2014) A1581}
  [\href{https://arxiv.org/abs/1303.1377}{{\ttfamily 1303.1377}}].

\bibitem{Alexandrou:2016izb}
C.~Alexandrou, S.~Bacchio, J.~Finkenrath, A.~Frommer, K.~Kahl and M.~Rottmann,
  \emph{{Adaptive aggregation-based domain decomposition multigrid for twisted
  mass fermions}},
  \href{https://doi.org/10.1103/PhysRevD.94.114509}{\emph{Phys. Rev. D}
  {\bfseries 94} (2016) 114509}
  [\href{https://arxiv.org/abs/1610.02370}{{\ttfamily 1610.02370}}].

\bibitem{Alexandrou:2018wiv}
C.~Alexandrou, S.~Bacchio and J.~Finkenrath, \emph{{Multigrid approach in
  shifted linear systems for the non-degenerated twisted mass operator}},
  \href{https://doi.org/10.1016/j.cpc.2018.10.013}{\emph{Comput. Phys. Commun.}
  {\bfseries 236} (2019) 51}
  [\href{https://arxiv.org/abs/1805.09584}{{\ttfamily 1805.09584}}].

\bibitem{tmLQCDGPU}
B.~Kostrzewa, S.~Bacchio, J.~Finkenrath, M.~Garofalo, F.~Pittler, S.~Romiti and
  C.~Urbach, \emph{{Twisted mass ensemble generation on GPU machines}},
  {\emph{PoS} {\bfseries LATTICE2022} (2022) 340}
  [\href{https://arxiv.org/abs/2212.06635}{{\ttfamily 2212.06635}}].

\bibitem{Baron:2010bv}
R.~Baron et~al., \emph{{Light hadrons from lattice QCD with light (u,d),
  strange and charm dynamical quarks}},
  \href{https://doi.org/10.1007/JHEP06(2010)111}{\emph{JHEP} {\bfseries 06}
  (2010) 111} [\href{https://arxiv.org/abs/1004.5284}{{\ttfamily 1004.5284}}].

\bibitem{EuropeanTwistedMass:2010voq}
{\scshape ETM} collaboration, R.~Baron et~al., \emph{{Computing K and D meson
  masses with $N_{f} = 2+1+1$ twisted mass lattice QCD}},
  \href{https://doi.org/10.1016/j.cpc.2010.10.004}{\emph{Comput. Phys. Commun.}
  {\bfseries 182} (2011) 299}
  [\href{https://arxiv.org/abs/1005.2042}{{\ttfamily 1005.2042}}].

\bibitem{ETM:2009ztk}
{\scshape ETM} collaboration, R.~Baron et~al., \emph{{Light meson physics from
  maximally twisted mass lattice QCD}},
  \href{https://doi.org/10.1007/JHEP08(2010)097}{\emph{JHEP} {\bfseries 08}
  (2010) 097} [\href{https://arxiv.org/abs/0911.5061}{{\ttfamily 0911.5061}}].

\bibitem{Alexandrou:2020okk}
C.~Alexandrou et~al., \emph{{Nucleon axial and pseudoscalar form factors from
  lattice QCD at the physical point}},
  \href{https://doi.org/10.1103/PhysRevD.103.034509}{\emph{Phys. Rev. D}
  {\bfseries 103} (2021) 034509}
  [\href{https://arxiv.org/abs/2011.13342}{{\ttfamily 2011.13342}}].

\bibitem{tmlqcd}
``{\texttt{tmLQCD} code}.'' \url{https://github.com/etmc/tmLQCD}.

\bibitem{Kotov:2021rah}
A.~Y. Kotov, M.~P. Lombardo and A.~Trunin, \emph{{QCD transition at the
  physical point, and its scaling window from twisted mass Wilson fermions}},
  \href{https://doi.org/10.1016/j.physletb.2021.136749}{\emph{Phys. Lett. B}
  {\bfseries 823} (2021) 136749}
  [\href{https://arxiv.org/abs/2105.09842}{{\ttfamily 2105.09842}}].

\bibitem{Ishikawa:2018jee}
{\scshape PACS} collaboration, K.-I. Ishikawa, N.~Ishizuka, Y.~Kuramashi,
  Y.~Nakamura, Y.~Namekawa, Y.~Taniguchi, N.~Ukita, T.~Yamazaki and T.~Yoshie,
  \emph{{Finite size effect on pseudoscalar meson sector in $2+1$ flavor QCD at
  the physical point}},
  \href{https://doi.org/10.1103/PhysRevD.99.014504}{\emph{Phys. Rev. D}
  {\bfseries 99} (2019) 014504}
  [\href{https://arxiv.org/abs/1807.06237}{{\ttfamily 1807.06237}}].

\bibitem{Shintani:2019wai}
{\scshape PACS} collaboration, E.~Shintani and Y.~Kuramashi, \emph{{Hadronic
  vacuum polarization contribution to the muon $g-2$ with $2+1$ flavor lattice
  QCD on a larger than (10 fm$)^4$ lattice at the physical point}},
  \href{https://doi.org/10.1103/PhysRevD.100.034517}{\emph{Phys. Rev. D}
  {\bfseries 100} (2019) 034517}
  [\href{https://arxiv.org/abs/1902.00885}{{\ttfamily 1902.00885}}].

\bibitem{Luscher:2003vf}
M.~Lüscher, \emph{{Lattice QCD and the Schwarz alternating procedure}},
  \href{https://doi.org/10.1088/1126-6708/2003/05/052}{\emph{JHEP} {\bfseries
  05} (2003) 052} [\href{https://arxiv.org/abs/hep-lat/0304007}{{\ttfamily
  hep-lat/0304007}}].

\bibitem{Clark:2006fx}
M.~A. Clark and A.~D. Kennedy, \emph{{Accelerating dynamical fermion
  computations using the rational hybrid Monte Carlo (RHMC) algorithm with
  multiple pseudofermion fields}},
  \href{https://doi.org/10.1103/PhysRevLett.98.051601}{\emph{Phys. Rev. Lett.}
  {\bfseries 98} (2007) 051601}
  [\href{https://arxiv.org/abs/hep-lat/0608015}{{\ttfamily hep-lat/0608015}}].

\bibitem{PACS-CS:2008bkb}
{\scshape PACS-CS} collaboration, S.~Aoki et~al., \emph{{$2+1$ Flavor lattice
  QCD toward the physical point}},
  \href{https://doi.org/10.1103/PhysRevD.79.034503}{\emph{Phys. Rev. D}
  {\bfseries 79} (2009) 034503}
  [\href{https://arxiv.org/abs/0807.1661}{{\ttfamily 0807.1661}}].

\bibitem{RBC:2014ntl}
{\scshape RBC-UKQCD} collaboration, T.~Blum et~al., \emph{{Domain wall QCD with
  physical quark masses}},
  \href{https://doi.org/10.1103/PhysRevD.93.074505}{\emph{Phys. Rev. D}
  {\bfseries 93} (2016) 074505}
  [\href{https://arxiv.org/abs/1411.7017}{{\ttfamily 1411.7017}}].

\bibitem{RBC:2020kdj}
{\scshape RBC-UKQCD} collaboration, R.~Abbott et~al., \emph{{Direct CP
  violation and the $\Delta I=1/2$ rule in $K\to\pi\pi$ decay from the standard
  model}}, \href{https://doi.org/10.1103/PhysRevD.102.054509}{\emph{Phys. Rev.
  D} {\bfseries 102} (2020) 054509}
  [\href{https://arxiv.org/abs/2004.09440}{{\ttfamily 2004.09440}}].

\bibitem{SIMULATeQCD}
``{\texttt{SIMULATeQCD}} public code repository.''
  \url{https://github.com/LatticeQCD/SIMULATeQCD}.

\bibitem{Bollweg:2021cvl}
D.~Bollweg, L.~Altenkort, D.~A. Clarke, O.~Kaczmarek, L.~Mazur, C.~Schmidt,
  P.~Scior and H.-T. Shu, \emph{{HotQCD on multi-GPU Systems}},
  \href{https://doi.org/10.22323/1.396.0196}{\emph{PoS} {\bfseries LATTICE2021}
  (2022) 196} [\href{https://arxiv.org/abs/2111.10354}{{\ttfamily
  2111.10354}}].

\bibitem{Bazavov:2020bjn}
{\scshape HotQCD} collaboration, A.~Bazavov et~al., \emph{{Skewness, kurtosis,
  and the fifth and sixth order cumulants of net baryon-number distributions
  from lattice QCD confront high-statistics STAR data}},
  \href{https://doi.org/10.1103/PhysRevD.101.074502}{\emph{Phys. Rev. D}
  {\bfseries 101} (2020) 074502}
  [\href{https://arxiv.org/abs/2001.08530}{{\ttfamily 2001.08530}}].

\bibitem{Bollweg:2021vqf}
{\scshape HotQCD} collaboration, D.~Bollweg et~al., \emph{{Second order
  cumulants of conserved charge fluctuations revisited: Vanishing chemical
  potentials}}, \href{https://doi.org/10.1103/PhysRevD.104.074512}{\emph{Phys.
  Rev. D} {\bfseries 104} (2021) }
  [\href{https://arxiv.org/abs/2107.10011}{{\ttfamily 2107.10011}}].

\bibitem{Bollweg:2022rps}
{\scshape HotQCD} collaboration, D.~Bollweg et~al., \emph{{Taylor expansions
  and Pad\'e approximants for cumulants of conserved charge fluctuations at
  nonvanishing chemical potentials}},
  \href{https://doi.org/10.1103/PhysRevD.105.074511}{\emph{Phys. Rev. D}
  {\bfseries 105} (2022) 074511}
  [\href{https://arxiv.org/abs/2202.09184}{{\ttfamily 2202.09184}}].

\bibitem{HotQCD:2019xnw}
{\scshape HotQCD} collaboration, H.~T. Ding et~al., \emph{{Chiral Phase
  Transition Temperature in $2+1$-Flavor QCD}},
  \href{https://doi.org/10.1103/PhysRevLett.123.062002}{\emph{Phys. Rev. Lett.}
  {\bfseries 123} (2019) 062002}
  [\href{https://arxiv.org/abs/1903.04801}{{\ttfamily 1903.04801}}].

\bibitem{Bruno:2014jqa}
{\scshape CLS} collaboration, M.~Bruno et~al., \emph{{Simulation of QCD with
  N$_{f} =$ 2 $+$ 1 flavors of non-perturbatively improved Wilson fermions}},
  \href{https://doi.org/10.1007/JHEP02(2015)043}{\emph{JHEP} {\bfseries 02}
  (2015) 043} [\href{https://arxiv.org/abs/1411.3982}{{\ttfamily 1411.3982}}].

\bibitem{Mohler:2017wnb}
D.~Mohler, S.~Schaefer and J.~Simeth, \emph{{CLS $2+1$ flavor simulations at
  physical light- and strange-quark masses}},
  \href{https://doi.org/10.1051/epjconf/201817502010}{\emph{EPJ Web Conf.}
  {\bfseries 175} (2018) 02010}
  [\href{https://arxiv.org/abs/1712.04884}{{\ttfamily 1712.04884}}].

\bibitem{RQCD:2022xux}
{\scshape RQCD} collaboration, G.~S. Bali et~al., \emph{{Scale setting and the
  light baryon spectrum in $N_f=2+1$ QCD with Wilson fermions}},
  \href{https://arxiv.org/abs/2211.03744}{{\ttfamily 2211.03744}}.

\bibitem{Mohler:2020txx}
D.~Mohler and S.~Schaefer, \emph{{Remarks on strange-quark simulations with
  Wilson fermions}},
  \href{https://doi.org/10.1103/PhysRevD.102.074506}{\emph{Phys. Rev. D}
  {\bfseries 102} (2020) 074506}
  [\href{https://arxiv.org/abs/2003.13359}{{\ttfamily 2003.13359}}].

\bibitem{cls}
``{Status of CLS configurations for $N_f=2+1$ flavours}.''
  \url{https://www-zeuthen.desy.de/alpha/public-cls-nf21/}.
\end{thebibliography}
\end{document}